\documentclass[final,3p,times,twocolumn]{elsarticle/elsarticle}

\usepackage{amssymb}
\usepackage{siunitx}
\usepackage{url}

\usepackage{subcaption}
\usepackage[font=footnotesize,labelfont=bf]{caption}

\usepackage{lineno}

\journal{Nuclear Physics A}

\begin{document}

\begin{frontmatter}



\title{Commissioning, Characterisation and Temperature Stabilisation of a 22000 Channel SiPM-on-Tile Hadron Calorimeter System}


\author{D. Heuchel$^{a}$ for the CALICE Collaboration}

\affiliation{organization={Deutsches Elektronen-Synchrotron (DESY)},
            addressline={Notkestr. 85}, 
            city={Hamburg},
            postcode={22607}, 
            country={Germany}}

\begin{abstract}
With the successful construction and operation of a highly granular hadron calorimeter system, featuring $\sim$22000  individually read out SiPM-on-tile channels, the CALICE collaboration has set the next milestone in proving the scalability of the concept for a future high energy linear collider experiment. For this large sample of photosensors a new approach of quality control was required to sufficiently characterise and monitor device parameters for both, test bench and in-situ beam test data. In the presence of temperature fluctuations during operation, it was possible to stabilise the SiPM responses with a fully automated adjustment of the bias voltage based on frequent temperature measurements, thanks to the excellent parameter uniformity of the devices. This contribution presents the results of SiPM parameter studies during the construction and commissioning phase and reports about the system performance and the experience of automated temperature compensation at system level during operation.
\end{abstract}


\end{frontmatter}


\section{Introduction}
\label{sec:introduction}
After the first successful demonstrations of the use of SiPMs in scintillator-based calorimeter systems over the last years, the CALICE collaboration has proven the scalability of the concept for large collider detectors with millions of channels \cite{Felix_Frank, ahcal_physics}. One key element in this development is the embedded electronic read-out  layers including surface mounted SiPMs, allowing for highly granular calorimeter layers with a compact front-end readout design. The latest Analogue Hadron Calorimeter (AHCAL) technological prototype features a total of $\sim$22000 SiPM-on-tile channels \cite{AHCAL_paper}. For this large amount of channels an extensive commissioning and characterisation campaign was required. Based on the determined excellent intrinsic SiPM parameter uniformity, a fully automated temperature compensation scheme for the bias voltage of the SiPMs has been implemented. During various beam test campaigns at the CERN SPS in 2018 this scheme allowed for an uniform and stable detector operation and performance.


\section{SiPM-on-Tile Technology}
\label{sec:sipm_tile}
\begin{figure}[!ht]
   \centering
   \hspace{-0.6cm}
   \begin{subfigure}[!ht]{0.22\textwidth}
   \centering
   \includegraphics[height=0.9\textwidth]{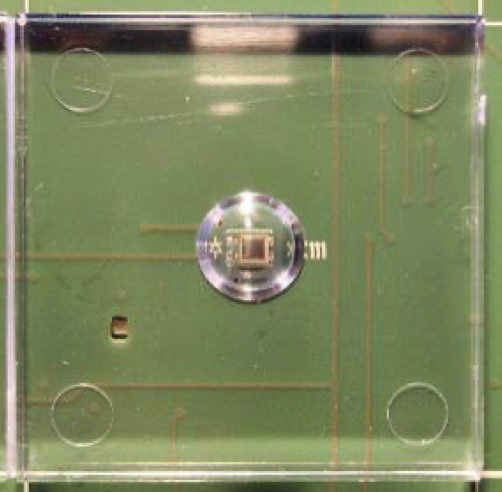}
   \caption{}
   \label{fig:sipm_on_tile} 
   \end{subfigure}
   \begin{subfigure}[!ht]{0.22\textwidth}
   \centering
   \includegraphics[height=0.9\textwidth]{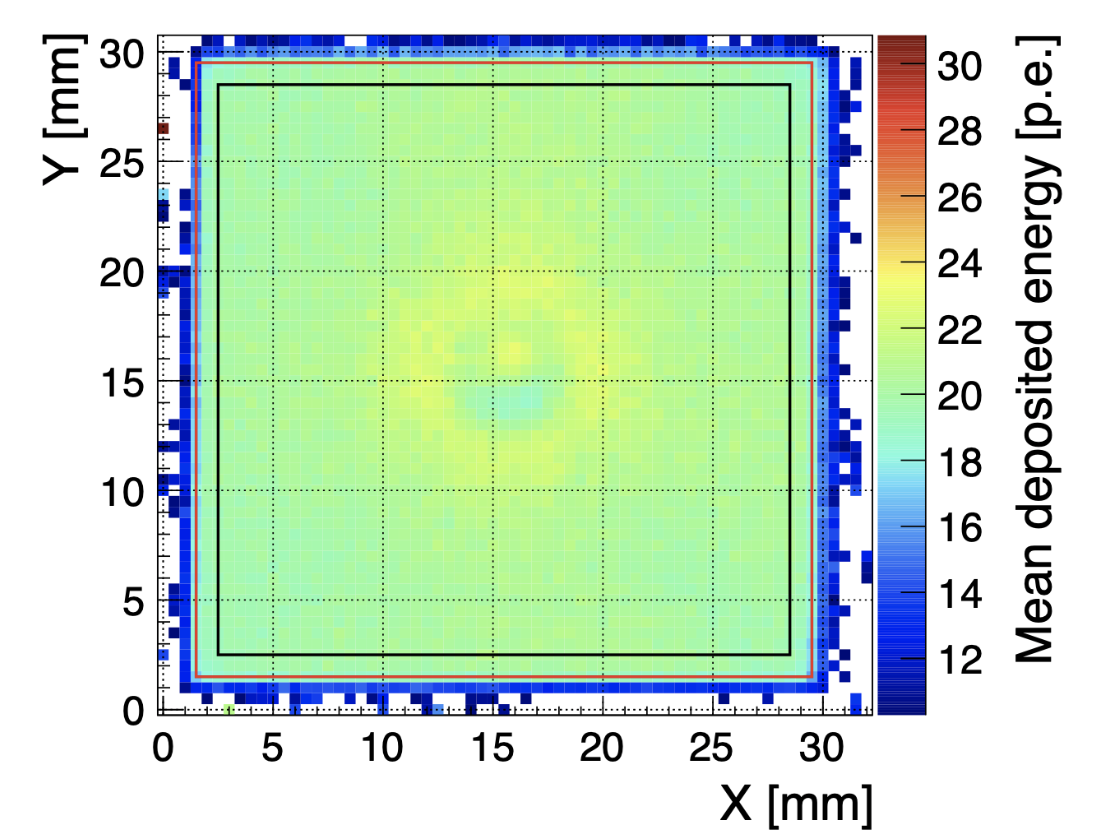}
   \caption{}
   \label{fig:tile_uniformity_data}
   \end{subfigure}
   \hspace{0.09cm}

\caption{\textbf{(a)} SiPM-on-tile calorimeter channel. \textbf{(b)} Spatial response uniformity of the SiPM-on-tile channel determined by a position scan with collimated electrons of a $^{90}$Sr source \cite{Liu}.}
\label{fig:sipm_tile_uniformity}
\end{figure}

Each channel of the hadron calorimeter system consists of a surface mounted silicon photomultiplier (SiPM) coupled to an injection-moulded polystyrene scintillator tile, as depicted in Fig. \ref{fig:sipm_on_tile}. The assembled type of SiPM (Hamamatsu MPPC S13360-1325PE) features 2668 pixels with a pixel pitch of \SI{25}{\micro \meter} covering an active are of \SI{1.3x1.3}{\milli \meter}. The device stands out in terms of its high photon detection efficiency (\SI{\sim 25}{\percent}), low dark count rate (typically \SI{\sim 70}{}-\SI{80}{\kilo \hertz}) and low cross-talk probability (\SI{\sim 1}{\percent}) compared to previous models. 

The scintillating tile, packaged in ESR foil with a reflectivity larger than \SI{98}{\percent} for the scintillator's spectral range, features dimensions of \SI{30x30x3}{\milli \meter} with a centralised dimple in which the SiPM is placed. The geometry of the spherical cavity is optimised for efficient and uniform light collection to the SiPM \cite{Liu}, as illustrated in Fig. \ref{fig:tile_uniformity_data}.

\section{Commissioning Phase and SiPM Mass Characterisation}
\label{sec:commissioning}
\begin{figure}[htp!]
\centering
\includegraphics[width=0.46\textwidth]{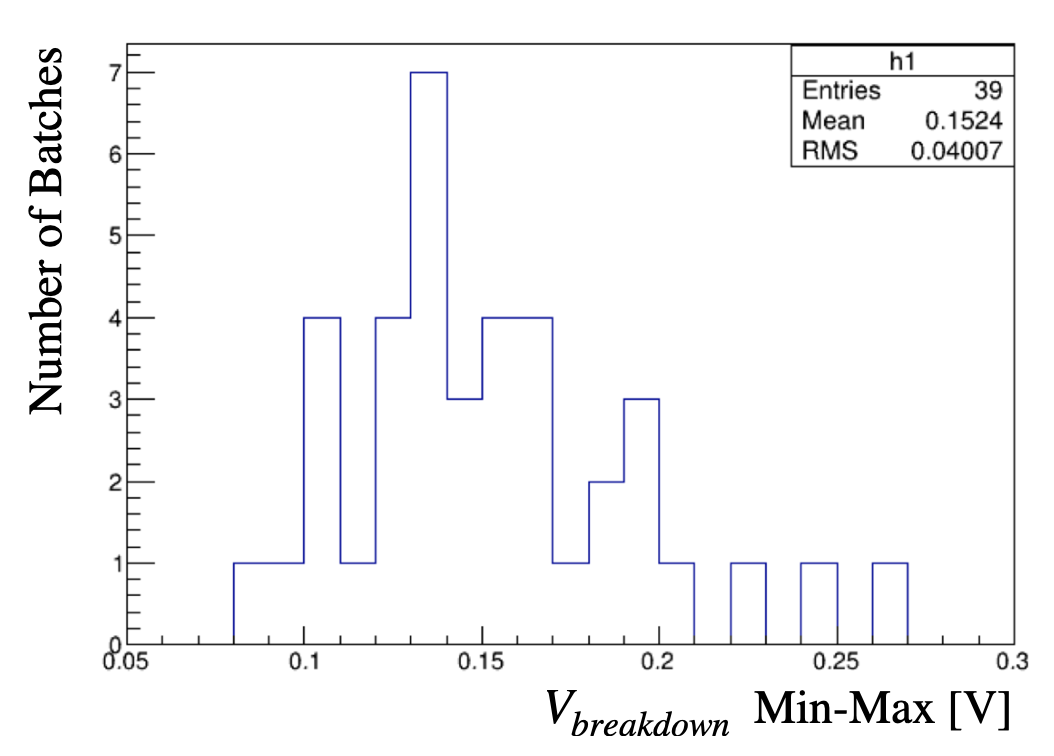}
\caption{Test bench measurement of largest $V_{breakdown}$ difference for SiPMs within a batch \cite{heidelberg}.}
\label{fig:v_break_min_max_variation}
\end{figure}

\begin{figure}[htp!]
\centering
\includegraphics[width=0.46 \textwidth]{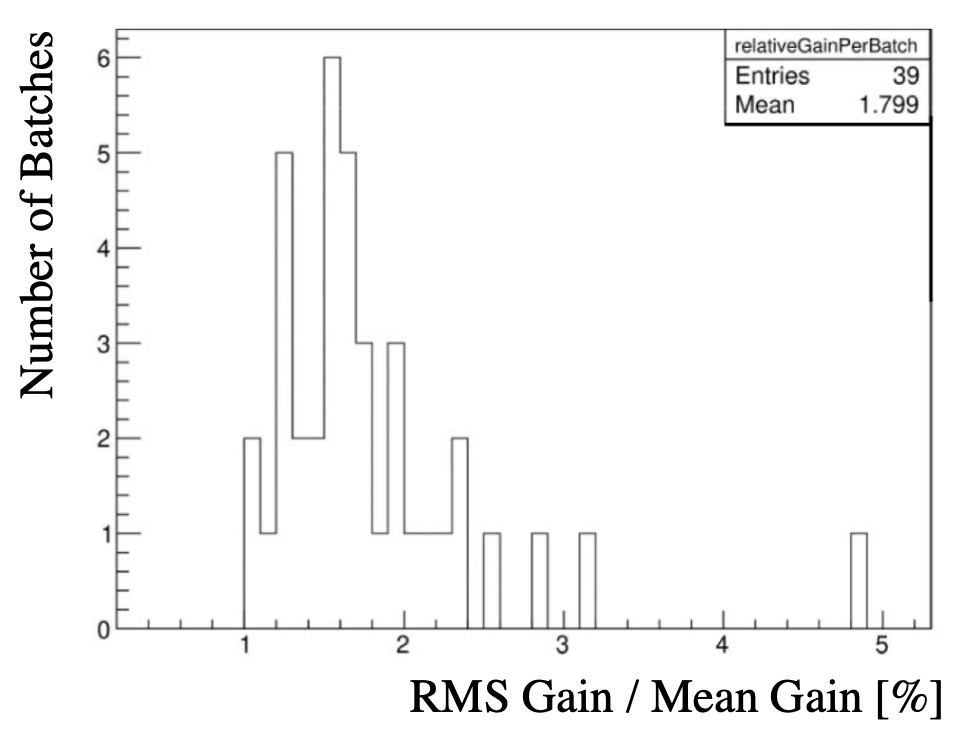}
\caption{Test bench measurement of gain variation for SiPMs within a batch.}
\label{fig:gain_variation}
\end{figure}

During the construction and commissioning phase of the calorimeter system in 2017/2018 the individual components were extensively tested and characterised before being assembled. In total 39 batches with 600 SiPMs each were delivered. Each batch was ordered with a required breakdown voltage uniformity of \SI{\pm 100}{\milli \volt}. For SiPM quality assurance and in order to prove the outstanding parameter homogeneity within the SiPM batches a dedicated test bench setup at the university of Heidelberg has been used \cite{heidelberg}. In a spot test procedure 24 SiPMs per batch were extensively characterised on the test bench, while the remaining 576 SiPMs were used for one calorimeter layer. In a first step, basic parameters for all spot test SiPMs were measured, which were well satisfying all defined requirements like a dark count rate lower than \SI{500}{\kilo \hertz} or a gain higher than \SI{3d5}{}. By scanning the point of breakdown voltage for each spot test SiPM, a mean min-max variation of \SI{\sim 150}{\milli \volt} per batch was found, see Fig. \ref{fig:v_break_min_max_variation}. Four batches showed a larger min-max difference than \SI{\sim 200}{\milli \volt}, however, only caused by one outlier SiPM per batch. This excellent homogeneity is reflected by the measured mean gain uniformity of \SI{1.8}{\percent} for SiPMs of the same batch, as depicted in Fig. \ref{fig:gain_variation}, allowing to build homogenous calorimeter layers with SiPMs of the same batch without having to test and characterise each individual device.

\section{The CALICE Analog Hadron Calorimeter Technological Prototype}
\label{sec:ahcal}
\begin{figure}[htp]
\centering
\includegraphics[width=0.42\textwidth]{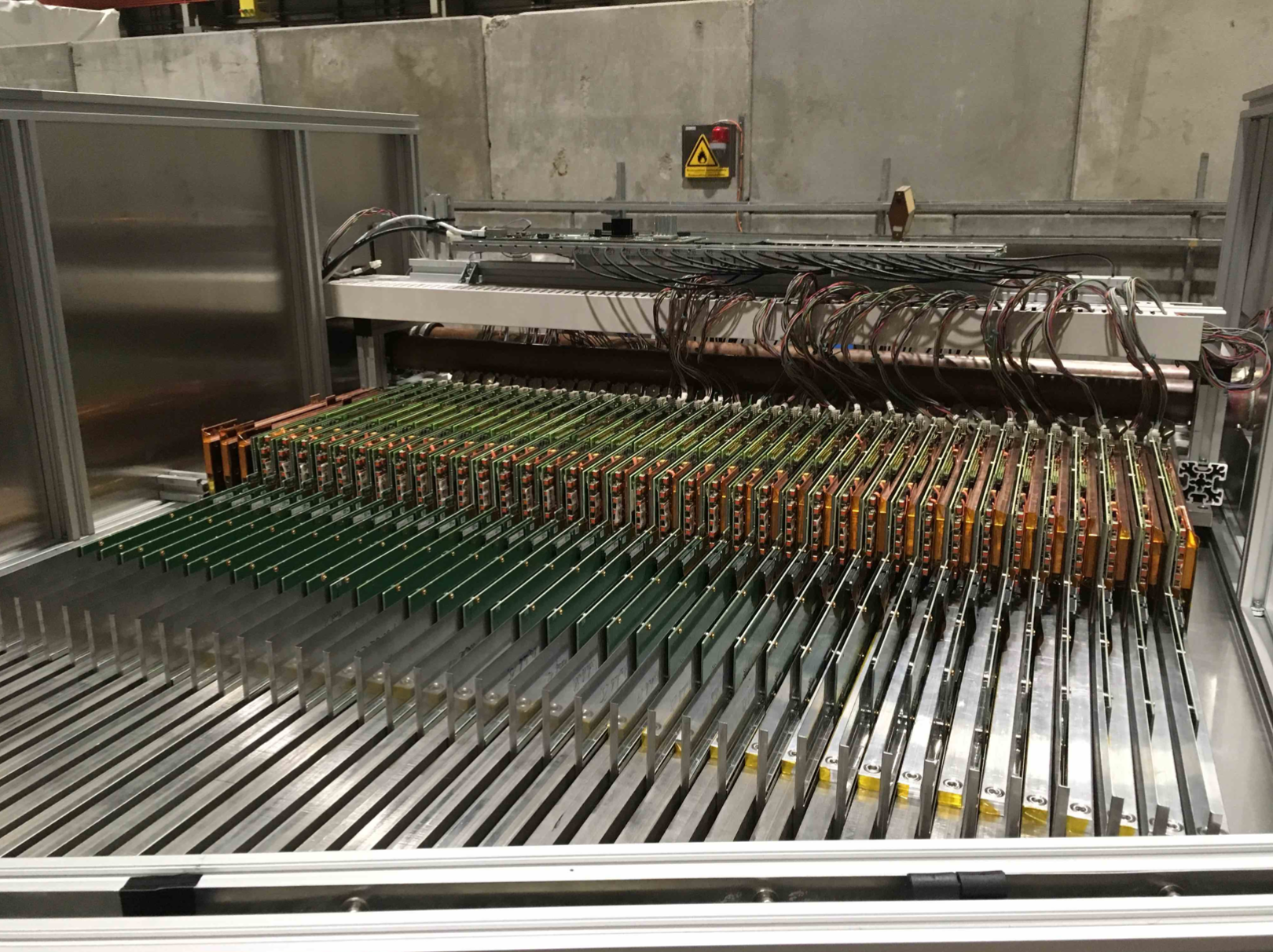}
\caption{Assembled AHCAL prototype featuring 38 active layers with a total of $\sim$22000 SiPM-on-tile channels.}
\label{fig:Prot}
\end{figure}

\begin{figure}[htp]
\centering
\includegraphics[width=0.4\textwidth]{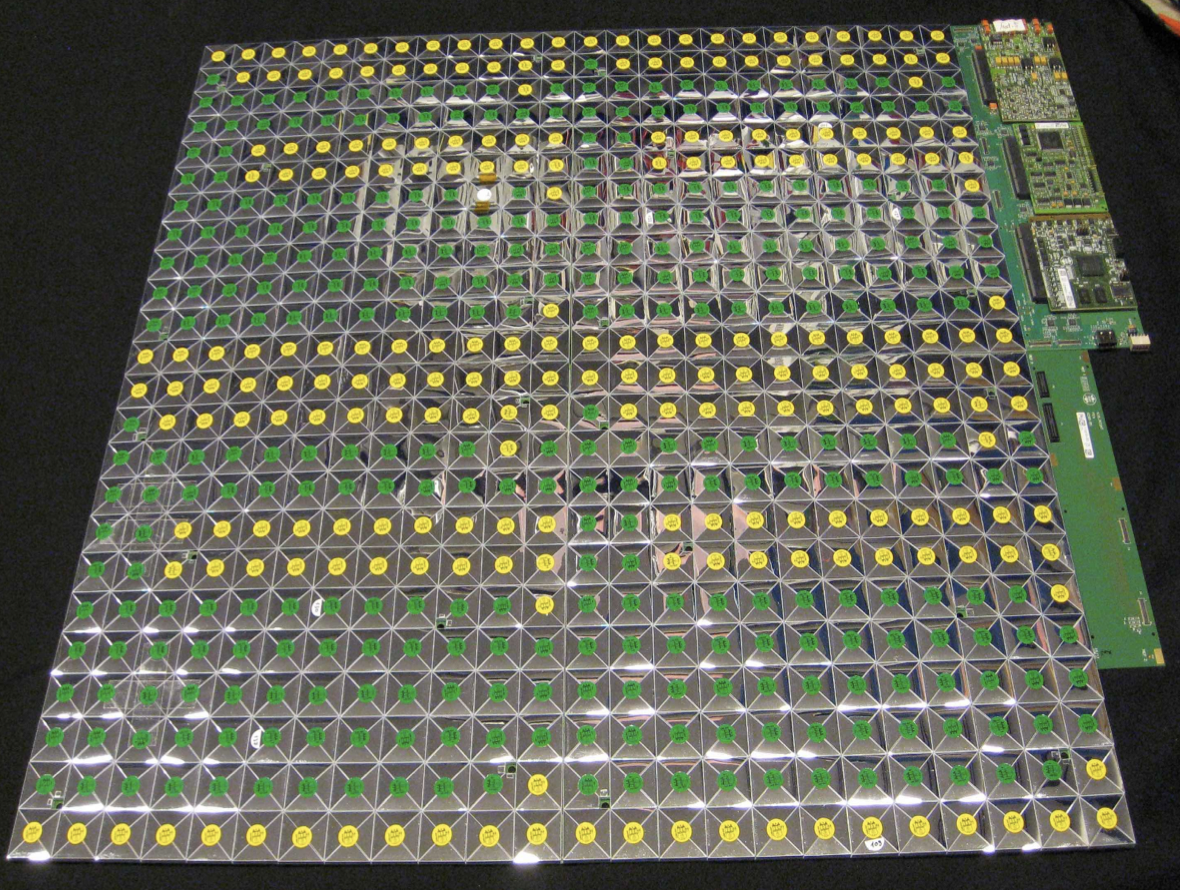}
\caption{AHCAL prototype layer consisting of 576 SiPM-on-tile channels covering an area of \SI{72x72}{\centi \meter}. Yellow and green mark different ESR foil layouts with respect to different LED positions for specific channels.}
\label{fig:ahcal_layer}
\end{figure}

The fully assembled AHCAL technological prototype consists of \SI{\sim 17.2}{\milli \meter} thick stainless steel absorber plates alternating with 38 active layers which are placed in the absorber structure as shown in Fig. \ref{fig:Prot}. In total it features 21888 SiPM-on-tile channels. Each layer (Fig. \ref{fig:ahcal_layer}) consists of 576 SiPMs operated at the same bias voltage and read out by 16 ASICs, which allows for an internal amplification and digitisation of the SiPMs analog signals \cite{bouchel}. 

The integration into the active detector volume and the capability to operate in power-pulsed mode for a reduced power consumption are key features of the compact front-end readout electronics \cite{mathias_front}. Additionally, temperature sensors are implemented in each layer and the detector has an integrated low intensity and short-pulsed LED system for the channel-wise calibration of the SiPM gain by measuring the single photon spectrum in-situ. During extensive beam test campaigns at SPS CERN and DESY, the detector system has proven stable and well controlled operation and data acquisition. Tens of millions of single particle events (muons, electrons, pions) have been recorded for sophisticated physics analyses, like the details of the particle shower development in the calorimeter. Due to the large number of channels in transversal and longitudinal direction these details can be resolved appropriately demonstrating the imaging capabilities of the system.


%

\section{On-Detector SiPM Readout and Gain Calibration}
\label{sec:on_detector_operation}
\begin{figure}[htp]
\centering
\includegraphics[width=0.4\textwidth]{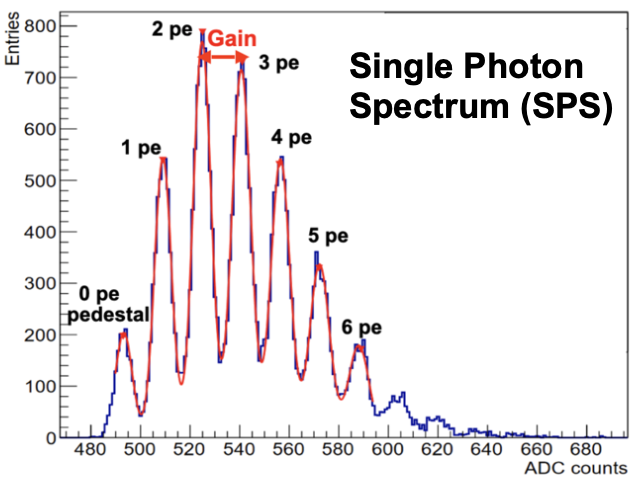}
\caption{Single photon spectrum for one AHCAL channel acquired in the LED calibration. The blue curve corresponds to the measured data while the red curve represents a multi-Gaussian fit to determine the gain.}
\label{fig:sps}
\end{figure}

\begin{figure}[htp]
\centering
\includegraphics[width=0.4\textwidth]{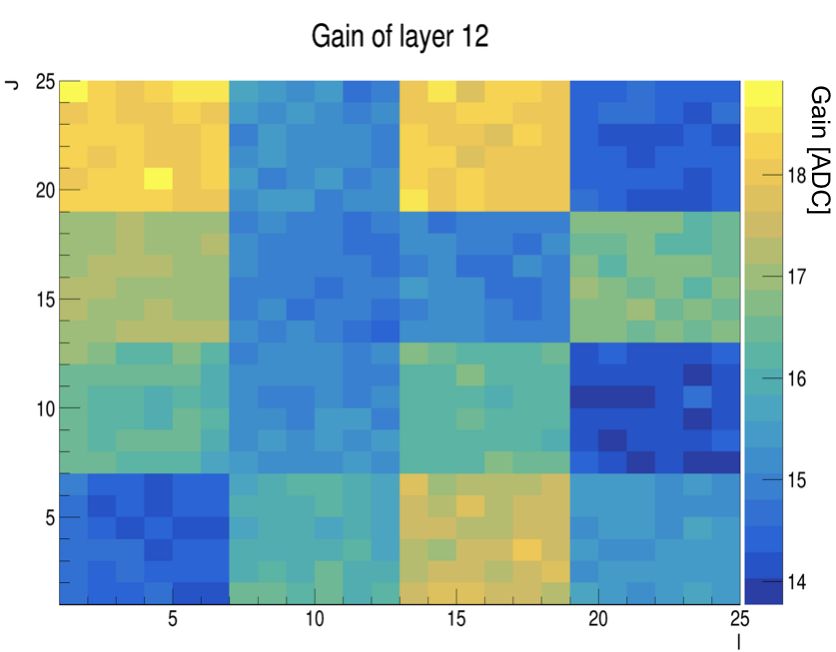}
\caption{Gain uniformity for one calorimeter layer. Each of the 16 squared areas with increased uniformity represents the 36 channels read out by the same ASIC.}
\label{fig:gain_map}
\end{figure}

In order to ensure system stability and to monitor the performance of all individual SiPMs, a daily in-situ LED gain calibration has been performed during beam test operation. The calibration procedure aims to determine the gain of each SiPM by measuring the distance between the equidistant single photon peaks in the single photon spectrum, as demonstrated in Fig. \ref{fig:sps}. The analog output signals of the SIPMs are amplified as well as digitised in the ASIC readout chips. Therefore, for monitoring purposes the gain is expressed in ADC (analogue-to-digital-counts) units, representing the intrinsic SiPM gain folded with the amplification factor of the chip component and the subsequent digitisation. More than \SI{99.9}{\percent} of the channels are determined to be fully functional, typically operating at a signal-to-noise ratio of \SI{\sim53}{}. The mean gain of the full measurement chain corresponds to ${\sim}$16.6 ADC/pixel with a RMS spread of ${\sim}$1.0 ADC/pixel, which is about \SI{6}{\percent}, across all channels of the detector system. For channels of the same ASIC the spread in gain is only \SI{\sim 2.5}{\percent}, as illustrated in Fig. \ref{fig:gain_map}, demonstrating that the gain variation between ASICs contributes dominantly to the overall gain spread in contrast to the intrinsic SiPM gain variation. The presented results validate not only a stable and reliable system operation, but also an excellent gain uniformity on readout level prior to the channel-wise calibration.

\section{Temperature Compensation and Gain Stability}
\label{sec:t_compensation}
\begin{figure}[htp]
\centering
\includegraphics[width=0.49\textwidth]{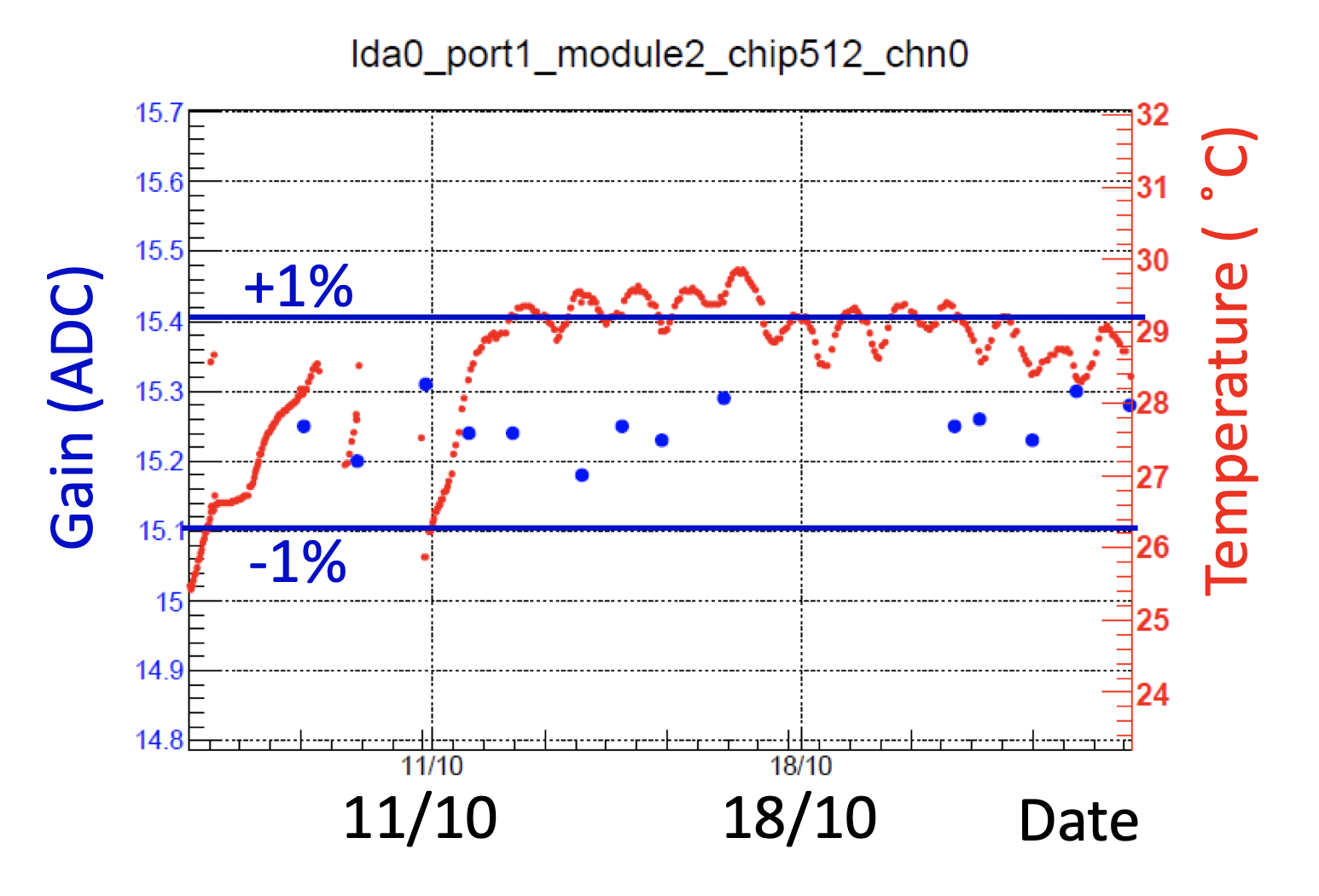}
\caption{Gain and temperature measurements for one AHCAL channel over the beam test period 6th - 24th October 2018.}
\label{fig:gain_vs_t}
\end{figure}

Most parameters of a SiPM, like the gain, are strongly dependent on the SiPM over-voltage:
\begin{equation}
\hspace{1.1cm} \Delta V = V_{bias} - V_{breakdown}(T).
\end{equation}
Since the breakdown voltage of a SiPM depends on the temperature (for the employed SiPMs $dV_{breakdown}/dT \sim \SI{54}{\milli \volt \per \kelvin}$ with a RMS spread of \SI{\sim 4}{\percent}), one should adjust the bias voltage according to temperature variations to keep the over-voltage (typically $\Delta V \approx  \SI{5}{\volt}$) constant. Therefore, the front-end power boards of the active calorimeter layers offer the function to adjust the bias voltage of the SiPMs, capable of coping with temperature differences of up to $\sim$\SI{40}{\kelvin}. Due to the excellent uniformity of the SiPM breakdown voltage, a common compensation for all SiPMs of a calorimeter layer is sufficient. According to temperature changes during a beam test period, the bias voltage is adjusted including a hysteresis for an increased stability to smaller fluctuations. The fully automated regular temperature compensation was used throughout the 2018 beam test periods. A gain stability within \SI{1}{\percent} for all SiPMs was achieved, as illustrated by Fig. \ref{fig:gain_vs_t}, which is within the precision of the gain extraction procedure.

\section{Summary and Outlook}
\label{sec:summary_outlook}
With the successful commissioning, characterisation and demonstration of temperature stabilisation of a $\sim$22000 SiPM-on-tile hadron calorimeter system, the CALICE collaboration has proven a well controlled employment and operation of a vast number of SiPMs. A mass characterisation campaign of spot test samples prior to the assembly validated the excellent quality of the devices and the outstanding parameter uniformity within the same batches. Therefore, $576$ uncharacterised SiPMs of the same batch could be operated with the same bias voltage at a time in one calorimeter layer. Based on daily in-situ LED gain calibration, a stable and reliable system operation was observed and an excellent uniformity on readout level has been achieved prior to channel-wise calibration. During the beam test periods in 2018, the layer-wise automated temperature compensation procedure allowed to keep the gain of all SiPMs of the calorimeter system constant within \SI{1}{\percent}. Based on the success of this technological prototypes concept, the SiPM-on-tile technology will also be employed in part of the endcap upgrade of the CMS detector at the LHC, which will feature $\sim$240000 channels with different types of SiPMs and different dimensions of scintillating tiles \cite{cms_hgcal}.

\section{Aknowledgements}
\label{sec:aknowledgements}
I would like to thank the CALICE collaboration and specifically the AHCAL colleagues for providing their research results, allowing me to present these results and for their continuous help and support. This project has received funding from the European Union’s Horizon 2020 Research and Innovation programme under Grant Agreement no. 654168 and 101004761.





\end{document}